\documentclass[twocolumn,amsmath,amssymb,showpacs,prb]{revtex4}

\usepackage{graphicx} 	
\usepackage{bm} 		
\usepackage{amssymb}
\usepackage{amsmath}

\begin{document}

\title{Resonating valence bonds and mean-field $d$-wave superconductivity in graphite}

\author{Annica M. Black-Schaffer}
 \affiliation{Department of Applied Physics, Stanford University, Stanford, California 94305}
 \author{Sebastian Doniach}
 \affiliation{Departments of Physics and Applied Physics, Stanford University, Stanford, California 94305}

\date{\today}
       
\begin{abstract}
We investigate the possibility of inducing superconductivity in a graphite layer by electronic correlation effects. 
We use a phenomenological microscopic Hamiltonian which includes nearest neighbor hopping and an interaction term which explicitly favors nearest neighbor spin-singlets through the well-known resonance valence bond (RVB) character of planar organic molecules. 
Treating this Hamiltonian in mean-field theory, allowing for bond-dependent variation of the RVB order parameter, we show that both $s$- and $d$-wave superconducting states are possible. The $d$-wave solution belongs to a two-dimensional representation and breaks time reversal symmetry. At zero doping there exists a quantum critical point at the dimensionless coupling $J/t = 1.91$ and the $s$- and $d$-wave solutions are degenerate for low temperatures. At finite doping the $d$-wave solution has a significantly higher $T_c$ than the $s$-wave solution. 
By using density functional theory we show that the doping induced from sulfur absorption on a graphite layer is enough to cause an electronically driven $d$-wave superconductivity at graphite-sulfur interfaces.
We also discuss applying our results to the case of the intercalated graphites as well as the validity of a mean-field approach.  
\end{abstract}


\maketitle
%
%
\section{Introduction}
The quasi-2D nature of graphite and the linear dispersion with zero density of states a the Fermi level, $\epsilon_{\bf k} = \hbar v_f |{\bf k}|$, of a single graphite sheet, called graphene, have been known for a long time.\cite{Wallace46} More recently, several experimental results have provided evidence that all physical properties of graphite are still not yet fully understood. 
For example, traces of superconducting behavior have recently been reported in both graphite\cite{Kopelevich00, Kopelevich06} and graphite-sulfur composites\cite{Kopelevich06,Moehlecke04,daSilva01}. 
Highly oriented pyrolithic graphite (HOPG) has been shown to display either a partial superconducting or a ferromagnetic-like response to an applied magnetic field even at room temperature depending on heat treatment and aging.\cite{Kopelevich00, Kopelevich06} For graphite-sulfur composites a partially superconducting state has been demonstrated below a critical temperature as high as $T_c = 35$ K with a continuous cross-over to a ferromagnetic-like behavior with increased magnetic field or temperature.\cite{Kopelevich06,Moehlecke04,daSilva01} 
X-ray analysis of the graphite-sulfur composites has shown no significant change in either the graphite or the orthorhombic sulfur structure and, with the active superconducting volume estimated to be only $0.05 \%$, the superconductivity can be understood in form of superconducting islands at the graphite-sulfur interfaces.\cite{Kopelevich06,daSilva01,Moehlecke04} 
Large observed magnetic field anisotropy strongly suggests that the superconducting correlations are localized to the graphite planes in both graphite\cite{Kopelevich06} and graphite-sulfur composites\cite{Moehlecke04}.

In addition to reports on traces of superconductivity, both a bulk magnetic-field-driven metal-insulator transition (MIT) and a reentrant insulator-metal transition (IMT) have been measured in graphite at low fields.\cite{Kopelevich03, KopelevichMIT06} Theoretical analysis has suggested that the MIT in graphite is due to the magnetic field and/or electronic interactions opening up an excitonic gap in the linear spectrum, in close analogy with dynamical chiral symmetry breaking in relativistic theories of (2+1)-dimensional Dirac fermions.\cite{Khveshchenko01a, Khveshchenko01b, Gorbar02}  
%
%
Thermal conductivity data showing a kink at the MIT is in agreement with general predictions of this theory.\cite{Ulrich04} 
Alternatively, conventional multi-band magnetoresistance theory has been used to explain the MIT as due to the semimetallic nature of graphite by a multiparameter fit to experimental data.\cite{Du05} However, recent magnetoresistance data\cite{KopelevichMIT06} show a non Fermi-liquid behavior with respect to magnetic field in the insulating phase, thus questioning the applicability of Fermi liquid theory. 
Yet another alternative explanation to the MIT in graphite comes from the excellent fit\cite{Kopelevichbook03} to the two-parameter scaling analysis suggested by Das and Doniach \cite{Das01} for a Bose metal-insulator transition. This analysis suggests the existence of a non-superconducting state of Cooper pairs in the zero temperature limit. Intriguingly, the MIT and a suppression of the superconducting signal have been found at the same applied field in HOPG samples,\cite{Kopelevich06} indicating that superconducting correlations might be important for the analysis of the MIT. Also, the field-induced reentrant IMT has been proposed as being caused by Cooper pair formation in the regime of Landau level quantization.\cite{Kopelevich03}

%
%
%
%
With the recent experimental discovery of single layers of graphene,\cite{Novoselov04} a wide array of theoretical and experimental results have also followed on graphene. Although not directly connected to what we report here, we mention briefly Refs.\ \onlinecite{Heersche06, Shailos06, Titov06, Cuevas06, Uchoa06} which include measurements of proximity induced superconductivity in graphene.\cite{Heersche06, Shailos06}
%

Our aim in this article is to investigate the effects of electronic correlations in a graphite layer. We are especially interested in the possibility to achieve superconductivity since many recent experiments seem to point to local superconducting correlations in graphite and graphitic compounds. We will assume that the inter-planar van der Waals coupling in graphite only acts as a small perturbation to the in-plane effects and in this work exclusively treat only the 2D graphene sheet. 
The carbon atoms in a graphene layer sits in a honeycomb lattice and are $sp^2$ hybridized, leaving, in the undoped case, one $\pi$-electron per carbon atom as the valence electrons.   
A well-known model for electronic correlations in the half-filled honeycomb lattice is the Hubbard model with nearest-neighbor hopping. In the high-$U$ limit, the Hubbard Hamiltonian can be rewritten as a $t$-$J$ Hamiltonian where the effective interaction, with coupling constant $J = 4t^2/U$, is between nearest-neighbor spins, such as to favor resonating valence bond (RVB) spin-singlet correlations. Due to the large on-site repulsion the Hilbert space is reduced to not include doubly occupied sites. At finite doping, spin-singlet pairs in this model become charged superconducting Cooper pairs. 
Choy {\it et al.} \cite{Choy95} used slave-boson mean-field theory to study the superconducting transition, associated with a non-zero order parameter for the RVB spin-singlet, for the $t$-$J$ model in the honeycomb lattice. They found a qualitatively similar phase diagram as for the square lattice (see e.g.\ Ref.\ \onlinecite{Suzumura87}). 
However, with typical values of the hopping $t \approx 2.5$ eV and the on-site repulsion $U \approx 6$ eV for the $p_z$-orbitals in graphite,\cite{Tchougreeff92} 
 it is questionable to approach the effect of correlations in graphite from the high-$U$ limit. With neither a perturbative weak-coupling approach nor the strong-coupling transformation to the $t$-$J$ model possible, theoretical approaches to correlations in graphite become highly limited. 
Early treatments of $p\pi$-bonded planar organic molecules and solids such as graphite by e.g.\ Pauling \cite{Paulingbook} rested heavily on the idea of RVBs, where spin-singlet bonds are favored compared to polar configurations (double and single occupancy of the $p_z$-orbitals). For example, good estimates were achieved in this way for the C-C bond distance, cohesive energy, and some excited state properties. Baskaran \cite{Baskaran02} used this, half-century old concept, in 2002 to propose a phenomenological Hamiltonian for graphite, including nearest neighbor hopping and a two-body interaction term favoring nearest neighbor spin-singlet formation. The interaction term is identical to the RVB term in the $t$-$J$ model and the only difference between the two models is that the phenomenological model makes the assumption that double site occupancy is still included in the Hilbert space and thus keeps the original kinetic energy term. This puts a stronger emphasis on the kinetic term, which in graphite is highly non-negligible. But still, the RVB interaction term will effectively promote the occurrence of spin-singlet nearest neighbor correlations, which, for strong enough coupling $J$ (or finite doping), will cause a condensation to a superconducting state. 

%
%
The proposed model is in fact equivalent to the $t$-$J$-$U$ model with the on-site repulsion $U$ set to zero. This model has been suggested as a good candidate for an effective one band model of the cuprates,\cite{Daul00} where $d$-wave superconductivity is triggered by the $J$ interaction whereas the $U$ repulsion introduces charge fluctuations in the model. It has been shown that a non-zero, repulsive $U$-term in fact enhances the $d$-wave superconducting correlations on a square lattice.\cite{Daul00,Plekhanov03,Arrachea05} 
Therefore, under the assumption that spin-singlet nearest neighbor correlations exist in graphite, thus warranting the $J$-term in the Hamiltonian, we expect, by ignoring the on-site repulsion $U$, to use a model which might slightly underestimate the superconducting correlations.
%
%
%
%
%

In this article we will treat Baskaran's effective Hamiltonian in mean-field theory but will allow the order parameter to be directionally dependent, allowing for non-$s$-wave superconducting states. We will show that a $d$-wave superconducting state is significantly more favorable than a $s$-wave state and that only small amounts of doping of the graphite layer is enough to initiate the onset of superconductivity. We will use this result to explore the possibility of $d$-wave superconductivity in pure graphite and graphite-sulfur composites and also in the intercalated graphites (GIC). 

The organization of the paper is as follows. In Section II we derive and numerically solve the BCS self-consistent equations for the spin-singlet pairing order parameter at a specified doping.  We deduce all possible symmetries of the order parameter as well as briefly discuss their properties. In Section III we report band structure calculations of graphite-sulfur systems indicating that sulfur can induce a hole doping sufficient to cause superconductivity in the graphite layer. In Section IV we study the intercalated graphites and show that the spin-singlet $d$-wave pairing in the $\pi$-band does not mix with the presumed phonon-pairing in the metallic band.  We point to different scenarios which would indicate why the GICs are not high-$T_c$ $d$-wave superconductors despite the fact that the intercalation induces a high doping into the $\pi$-band. Finally, in Section V we briefly discuss the applicability of mean-field theory and summarize our results.
%
%
\section{RVB correlations in graphene}
\begin{figure}
\includegraphics[scale = 1]{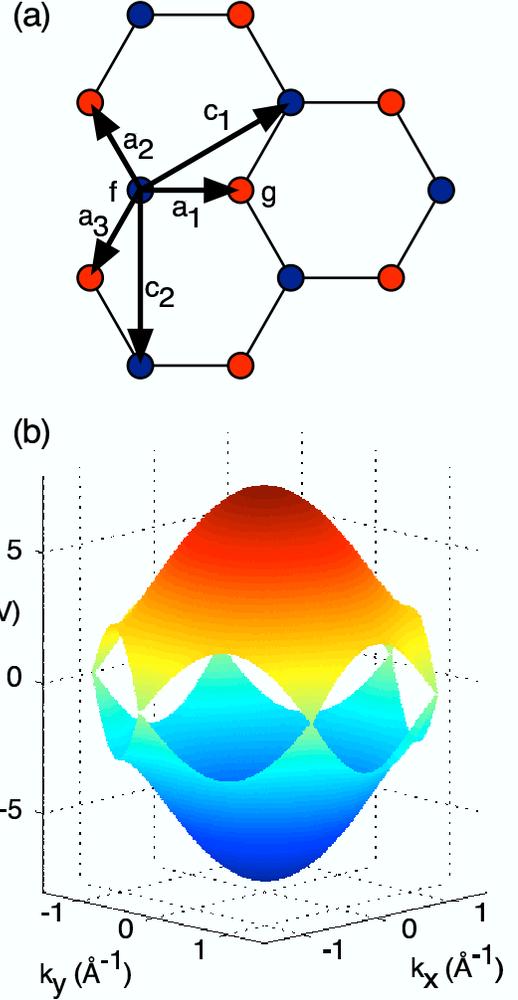}
\caption{\label{fig:graphene} (Color online) Part (a) shows the 2D honeycomb lattice with lattice vectors, ${\bf c}_1$ and ${\bf c}_2$, and nearest neighbor vectors,  ${\bf a}_1$, ${\bf a}_2$, and  ${\bf a}_3$, indicated as well as the two different lattice sites, $f$ (blue) and $g$ (red). (b) Band structure of graphene, $\pm t\epsilon_{\bf k}$, in the first Brillouin zone. The band structure close to the corners of the Brillouin zone has a linear dispersion with a point-like Fermi surface at zero doping.}
\end{figure}

Motivated by the well-recognized RVB spin-singlet character of a graphite layer we model a graphite layer using the following Hamiltonian: \cite{Baskaran02}
%
\begin{align}
\label{eq:H_eff}
H_{{\rm eff}} & =  -t\!\!\!\sum_{<i,j>,\sigma} \!\!\!(f_{i\sigma}^\dagger g_{j\sigma} + g_{i\sigma}^\dagger f_{j\sigma}) \nonumber \\&
+ \mu \sum_{i,\sigma} (f_{i\sigma}^\dagger f_{i\sigma} + g_{i\sigma}^\dagger g_{i\sigma})  - J\!\sum_{<i,j>}h_{ij}^\dagger h_{ij},
\end{align}
where $<\! i,j\!>$ indicates sum over nearest neighbors, $f_{i\sigma}^\dagger$ is the creation operator on the $A$ site of the honeycomb lattice, and $g_{i\sigma}^\dagger$ on the $B$ site, see Figure \ref{fig:graphene}(a). The nearest neighbor spin-singlet creation operator is
%
\begin{align}
\label{eq:h_ij}
h_{ij}^\dagger = \frac{1}{\sqrt{2}}(f_{i\uparrow}^\dagger g_{j\downarrow}^\dagger - f_{i\downarrow}^\dagger g_{j\uparrow}^\dagger)
\end{align}
when $i\in A$-site and has $f\leftrightarrow g$ for $i\in B$-site.

The energy gain from a nearest neighbor spin-singlet bond is given by
%
\begin{align}
\label{eq:RVBterm}
-Jh_{ij}^\dagger h_{ij} = J \left(  {\bf S}_i {\bf \cdot S}_j - \frac{1}{4}n_i n_j \right),
\end{align}
which clearly displays the similarity between our phenomenological model and the effective $t$-$J$ model for the high-$U$ limit of the Hubbard model. Baskaran\cite{Baskaran02} estimated the effective coupling by calculating the spin-singlet two-electron ground state energy to $J = \frac{1}{2}\sqrt{U^2+16t^2} - \frac{U}{2}$ which with $ t \approx 2.5$ eV and $U \approx 6$ eV in graphite gives $J/t \sim 1$. 

To estimate the superconducting transition temperature we use mean field theory. A complete Hartree-Fock-Bogoliubov factorization will yield order parameters not only for the spin-singlet pairing but also for the nearest neighbor hopping and site occupation number. Assuming the latter two to both be bond and site independent, we simplify the formalism by assuming that they are already included in the effective values of $t$ and $\mu$. Kotliar {\it et al.} \cite{Kotliar&Liu88}  showed that for the $t$-$J$ model on a square lattice this is indeed the case even though the superconducting order parameter was non-$s$-wave.
With this assumption the mean field Hamiltonian can be written in reciprocal space as
%
\begin{align}
\label{eq:HMF}
H_{{\rm MF}}  & = \sum_{{\bf k},{\bf a},\sigma}-t e^{i {\bf k\cdot a}} f_{{\bf k}\sigma}^\dagger g_{{\bf k}\sigma} + {\rm H.c.} \nonumber \\
& + \sum_{{\bf k},\sigma} \mu (f_{{\bf k}\sigma}^\dagger f_{{\bf k}\sigma} + g_{{\bf k}\sigma}^\dagger g_{{\bf k}\sigma}) \nonumber \\
& - \sum_{{\bf k},{\bf a}}\sqrt{2}J\Delta_{J{\bf a}}e^{i{\bf k\cdot a}}(f_{{\bf k}\uparrow}^\dagger g_{{\bf -k}\downarrow}^\dagger - f_{{\bf k}\downarrow}^\dagger g_{{\bf -k}\uparrow}^\dagger) + {\rm H.c.} \nonumber \\
& + NJ\sum_{\bf a} 4|\Delta_{J{\bf a}}|^2,
\end{align}
where $\Delta_{J{\bf a}} = \langle h_{i,i+{\bf a}}\rangle$ is the order parameter for RVB spin-singlet pairing. Here  ${\bf a} = \pm {\bf a}_1, {\bf a}_2$, or ${\bf a}_3$ denotes the three inequivalent directions for nearest neighbors in the honeycomb lattice, positive sign for $i \in {\rm A}$ and negative sign for $i \in {\rm B}$, see Figure \ref{fig:graphene}. The last constant term is irrelevant for the following discussion and will be ignored hereafter.
%
The kinetic energy term can be diagonalized by the following transformation: 
%
\begin{align}
\label{eq:banddiag}
\left( \begin{array}{c}
f_{{\bf k}\sigma}  \\ g_{{\bf k}\sigma}
\end{array} \right) = \frac{1}{\sqrt{2}}
\left( \begin{array}{c}
d_{{\bf k}\sigma} + c_{{\bf k}\sigma} \\ e^{-i\varphi_{\bf k}}(d_{{\bf k}\sigma} - c_{{\bf k}\sigma})
\end{array} \right).
\end{align}
Here $d_{{\bf k}\sigma}^\dagger$ creates an electron in the lower $\pi$-band, $c_{{\bf k}\sigma}^\dagger$ creates an electron in the upper $\pi$-band, and $\varphi_{\bf k} = \rm{arg}\left( \sum_{\bf a}e^{i{\bf k \cdot a}}\right)$. The $k$-dependence of the $\pi$-bands is given by $\epsilon_{\bf k} =|\sum_{\bf a}e^{i{\bf k \cdot a}}|$, see Figure \ref{fig:graphene}(b).
With this transformation the Hamiltonian (\ref{eq:HMF}) can be written as
%
\begin{align}
\label{eq:HMFk}
H_{{\rm MF}}  & = \sum_{{\bf k}\sigma} \left[ (\mu + t\epsilon_{\bf k})c_{{\bf k}\sigma}^\dagger c_{{\bf k}\sigma} + (\mu - t\epsilon_{\bf k})d_{{\bf k}\sigma}^\dagger d_{{\bf k}\sigma}  \right. \nonumber \\
& - \sqrt{2}J \sum_{\bf a} \Delta_{J{\bf a}} \cos({\bf k\cdot a}-\varphi_{\bf k})(d_{{\bf k}\uparrow}^\dagger d_{{\bf -k}\downarrow}^\dagger - c_{{\bf k}\uparrow}^\dagger c_{{\bf -k}\downarrow}^\dagger) \nonumber \\
& -  \sqrt{2} i J \sum_{\bf a} \left. \Delta_{J{\bf a}} \sin({\bf k\cdot a}-\varphi_{\bf k})(c_{{\bf k}\uparrow}^\dagger d_{{\bf -k}\downarrow}^\dagger - d_{{\bf k}\uparrow}^\dagger c_{{\bf -k}\downarrow}^\dagger) \right]. 
\end{align}
As seen in Eq.\ (\ref{eq:HMFk}), the RVB spin-singlet order parameter induces both inter- and intraband pairing in the two $\pi$ bands. The intraband pairing, second row in Eq. (\ref{eq:HMFk}), has an order parameter that is even in $k$-space and corresponds to a singlet pairing state. However, the interband coupling gives a triplet pairing with  the so called ${\bf d}$-vector being ${\bf d} = d_z({\bf k})\hat{{\bf z}}$ and with the order parameter odd in $k$-space. For a bond independent $s$-wave order parameter, the interband order parameter is identically zero, but this is not necessarily the case for an arbitrary wave symmetry.
The Hamiltonian (\ref{eq:HMFk}) can be diagonalized with a double Bogoliubov-Valatin transformation, using two sets of quasi-particle operators: 
%
\begin{align}
\label{eq:Bog}
H_{{\rm MF}}   = \sum_{\bf k} & \left[ E_{1,{\bf k}} ( \beta_{{\bf k}0}^\dagger \beta_{{\bf k}0} + \beta_{{\bf k}1}^\dagger \beta_{{\bf k}1} ) \right. \nonumber \\
& \left. + E_{2,{\bf k}} ( \gamma_{{\bf k}0}^\dagger \gamma_{{\bf k}0} + \gamma_{{\bf k}1}^\dagger \gamma_{{\bf k}1}) \right],
\end{align}
where $E_{i,{\bf k}}$ are the two distinct eigenvalues of $H_{\rm MF}$. 
Close to $T_c$ the gap parameters $\Delta_{J{\bf a}}$ should go to zero and we can expand the self-consistency equations in the gap parameters to finally arrive at the following self-consistency equations at the phase transition temperature:
\begin{widetext}
\begin{align}
\label{eq:scdelta}
\delta & = \frac{1}{2N}\sum_{\bf k} \tanh\left( \frac{\beta_c(t\epsilon_{\bf k}+\mu)}{2}\right) - \tanh\left( \frac{\beta_c(t\epsilon_{\bf k}-\mu)}{2}\right) \\
\label{eq:scDelta}
\Delta_{J{\bf a}}^\dagger & = \frac{J}{N}\sum_{\bf k} \sum_{\bf b} \left[ \cos({\bf k\cdot a}-\varphi_{\bf k}) \cos({\bf k\cdot b}-\varphi_{\bf k}) 
\left( \frac{\tanh(\beta_c (t\epsilon_{\bf k}+\mu)/2)}{2(t\epsilon_{\bf k} +\mu)} + \frac{\tanh( \beta_c(t\epsilon_{\bf k}-\mu)/2)}{2(t\epsilon_{\bf k}-\mu)} \right) \right. \nonumber \\
& \left. \phantom{mmmmmm}+ \sin({\bf k\cdot a}-\varphi_{\bf k})\sin({\bf k\cdot b}-\varphi_{\bf k}) \left( \frac{\sinh(\beta_c\mu)}{2\mu \cosh( \beta_c (t\epsilon_{\bf k}+\mu)/2) \cosh( \beta_c(t\epsilon_{\bf k}-\mu)/2)} \right) \right] \Delta_{J {\bf b}}^\dagger.
\end{align}
\end{widetext}
Here $N$ is the number of unit cells, $\beta_c = (k_BT_c)^{-1}$ is the inverse critical temperature, and $\delta$ is the doping away from half-filled $\pi$-bands in terms of holes per C atom. Note however that both self-consistency equations are invariant with respect to the sign of the doping, i.e.\ electron and hole doping give the same results.
The two first terms in Eq.\ (\ref{eq:scDelta}) come from the intraband coupling and have the familiar $\tanh(\beta E/2)/(2E)$ BCS-form. The last term is due to the triplet interband pairing. For finite doping or the reasonably low temperatures we are interested in, its contribution is minimal, and we will therefore often ignore this part.
For a $s$-wave solution, i.e. $\Delta_{J{\bf a}}$ independent of ${\bf a}$, Eq.\ (\ref{eq:scDelta}) reduces to the comparable  equation found in Ref.\ \onlinecite{Choy95}, where the $t$-$J$ model was applied to the honeycomb lattice. For a general solution we can, by using ${\bf \Omega} = (\Delta_{J{\bf a}_1}^\dagger,\Delta_{J{\bf a}_2}^\dagger,\Delta_{J{\bf a}_3}^\dagger)^T$, write Eq.\ (\ref{eq:scDelta}) in matrix form as 
%
\begin{align}
\label{eq:gapmatrixeq}
\frac{1}{J} {\bf \Omega} = 
\left( \begin{array}{ccc}
D & B & B \\ B & D & B \\ B & B & D \\
\end{array} \right) {\bf \Omega},
\end{align}
where $B=B(\beta_c)$ is the RHS of (\ref{eq:scDelta}) when ${\bf a} \neq {\bf b}$ and $D=D(\beta_c)$ is the RHS when ${\bf a} = {\bf b}$. The eigenvalues  to the above equation is easily found to be
%
\begin{align}
\label{eq:eigenvalues}
\frac{1}{J} =  \left\{
\begin{array}{ll}
D+2B, & {\rm extended} \ s{\rm-wave} \\
D-B, & d{\rm -wave} \ {\rm (and}\ p{\rm-wave}).
\end{array} 
\right.
\end{align}
The first solution, the extended $s$-wave, has eigenvector $\Delta_{J{\bf a}} = (1,1,1)$ which leaves the intraband order parameter in Eq.\ (\ref{eq:HMFk}) proportional to $\epsilon_{\bf k}$ and the interband order parameter identically equal to zero. This extended $s$-wave symmetry belongs naturally to $A_{1g}$, the identity irreducible representation, of the crystal point group $D_{6h}$.
The second solution is two-fold degenerate and spanned by vectors $\Delta_{J{\bf a}} = \{(2,-1,-1),(0,-1,1)\}$. It leaves both the intra- and interband order parameters non-zero. The intraband order parameter belongs to the $E_{2g}$ irreducible representation of $D_{6h}$ whereas the interband order parameter belongs to the $E_{1u}$ representation. Both of these representations are two-fold degenerate with $E_{2g}$ representing the two four-fold symmetries in a hexagonal crystal, $(k_x^2-k_y^2,k_x k_y)$, i.e.\ a $d$-wave state, whereas $E_{1u}$ has a simple odd symmetry, $(k_x,k_y)$, associated with a $p$-wave state. Note that for finite doping or low temperatures, the interband pairing is strongly suppressed and the system will then effectively have only a $d$-wave order parameter. Thus, we will hereafter simply call this solution $d$-wave.
%
%
\subsection{Numerical results}
\begin{figure}
\includegraphics[scale = 1]{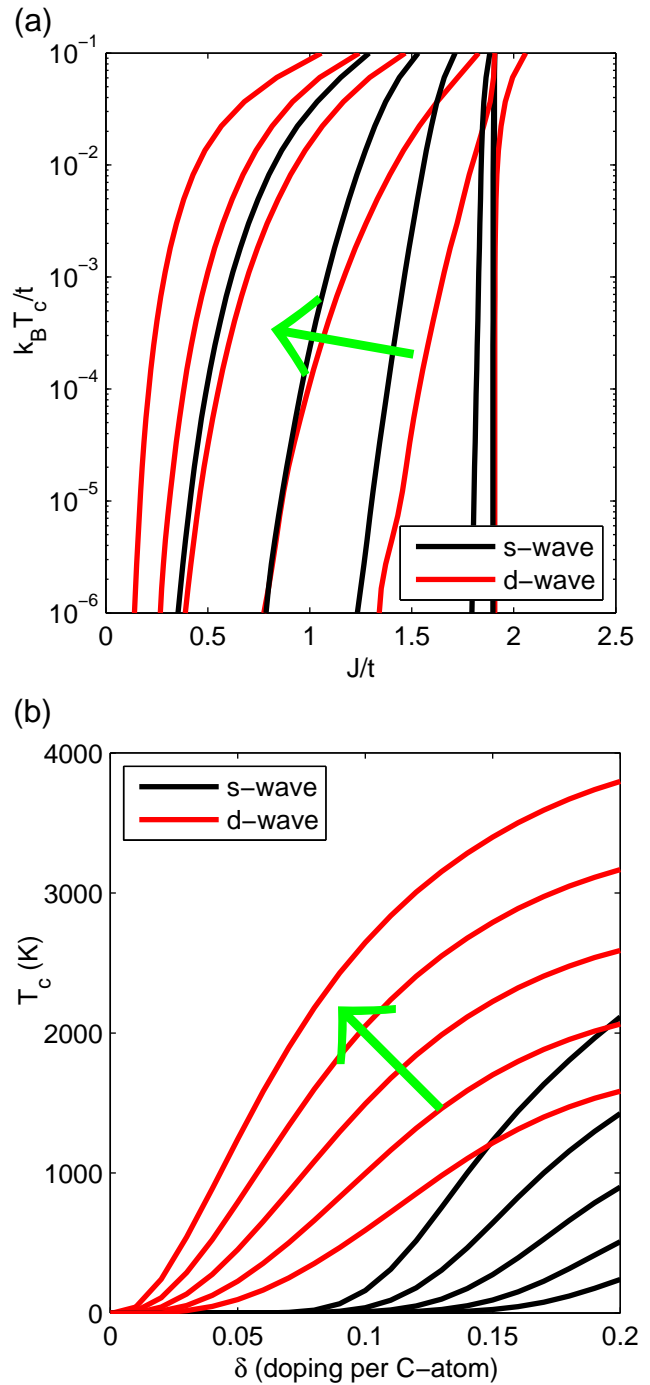}
\caption{\label{fig:TcvsJt} (Color online) (a) Dimensionless transition temperature $k_B T_c/t$ as function of $J/t$ for doping levels $\delta = 0, 0.001, 0.01, 0.05, 0.1,$ and 0.2. Big green arrow indicated direction of increasing doping. The dimensionless temperature scale spans approximately 0.03 K to 3000 K for $t = 2.5$ eV. (b) Transition temperature $T_c$ calculated using $t = 2.5$ eV  as a function of doping level $\delta$ for $J/t = 0.8,0.9,1,1.1,$ and 1.2. Big green arrow indicate direction of increasing $J/t$.}
\end{figure}

For given $t, J,$ and doping level $\delta$ (electron or hole doping) we can numerically solve Eqs.\ (\ref{eq:scdelta}) and (\ref{eq:scDelta}), by summing over the irreducible part of the first Brillouin zone, to get the critical temperature $T_c$ for both the $s$-wave and $d$-wave solutions. Figure \ref{fig:TcvsJt}(a) shows the dimensionless quantity $k_BT_c/t$ as a function of $J/t$ for a few different doping levels whereas Figure \ref{fig:TcvsJt}(b) shows $T_c$ as a function of doping level for $t = 2.5$ eV and different values of $J/t$ around the expected value of 1 for graphene.
%
As seen in Figure \ref{fig:TcvsJt}, for zero doping, the $s$-wave and $d$-wave solutions are degenerate for low temperatures with $J/t \geq J_c/t = 1.91$ necessary for a superconducting transition. This is indicative of a quantum critical point at $J/t = 1.91$ for $\delta = 0$. This behavior can be understood when considering the point-like Fermi surface in graphene at zero doping. 
Our value of $J_c/t$ differs from Ref.\ \onlinecite{Baskaran02} where the critical RVB coupling was estimated using a less detailed Cooper pair analysis to $J_c/t \approx 3$ at zero doping. However, with $J/t \sim 1$ in graphene the qualitative prediction of no superconducting state in graphene is unchanged. 

More important however, is that the $d$-wave solution will always be favored for $0< \delta < 0.4$ because then $B$ in Eq. (\ref{eq:gapmatrixeq}) is negative. In fact, for finite doping the $d$-wave solution has a much higher $T_c$ than the $s$-wave solution. While any linear combination of the two $d$-wave eigenvectors yields a valid solution at $T_c$, we found numerically that only  combinations that leave $E_{i{\bf k}}$ six-fold symmetric will be self-consistent solutions below $T_c$.  There is only one such combination, ${\bf \Omega}~=~ (1,\frac{-1+\sqrt{3}i}{2},\frac{-1-\sqrt{3}i}{2})^T$, up to an arbitrary factor and permutations among the bonds. 
  This real-space order parameter leads to complex-valued order parameters in reciprocal space and, thus, this solution is in fact time-reversal symmetry (TRS) breaking. The $d$-wave solution can be TRS breaking because it belongs to a multi-dimensional irreducible representation, in this case $E_{2g}$,  of the $D_{6h}$ group and, if not ignored, the $p$-wave solution breaks TRS because it is non-unitary. Kuznetsova {\it et al.} \cite{Kuznetsova05} recently pointed out that, in the weak-coupling limit, an order parameter belonging to $E_{2g}$ in a hexagonal crystal will be TRS breaking and ferromagnetic. In light of the experimental findings of a possible interaction between superconducting and ferromagnetic order parameters in graphite-sulfur composites,\cite{Moehlecke04,Kopelevich06} the TRS breaking property of the $d$- (and $p$)-wave is very intriguing. Note also that this $d$-wave solution only has a node at $\Gamma$ and thus the quasiparticle spectrum is nodeless.
The $d$-wave solution has the highest $T_c$ for $\delta = 0.25$. This is the doping at which the the Fermi level is located at the separatrix in the band structure, $t \epsilon_{\bf k}$, and thus the system has the largest possible Fermi surface. 

As seen in Figure \ref{fig:TcvsJt}(b) the superconducting state in graphene will only be realized for large enough doping levels if we assume $J/t \sim 1$. 
%
%
In recent experiments\cite{Novoselov04,Novoselov05,Zhang05} on a single graphene layer doping was induce in the graphene by means of applying a gate voltage. However, the largest achievable doping level was only $\delta \sim 0.008$ electrons/holes per C-atom which is too small for even the $d$-wave solution to have any sizable mean-field superconducting transition temperature.
%
%
%
The experimentally measured electron and hole concentrations in pure graphite, due to $c$-axis coupling, is even smaller, $\sim 10^{-4}$ electrons/holes per C atom. 
%
%
\section{Graphite-sulfur composites}
The results in the above section indicate that a TRS breaking $d$-wave superconducting state is possible in sufficiently electron or hole doped graphene/graphite. One such possible system might be the graphite-sulfur composites. With the graphite-sulfur composites retaining both graphite and sulfur x-ray signatures and the superconductivity possibly located only to the C-S interfaces\cite{Kopelevich06,Moehlecke04,daSilva01} we suggest that it might be a consequence of hole doping into the interface graphite layers, which then would induce a $d$-wave superconductivity locally at the interface.

In order to deduce a possible charge transfer between C and S we have preformed density functional theory (DFT) atomic relaxation and band structure calculations of various graphite-sulfur interface systems. Solid sulfur is a molecular solid consisting of S$_8$ rings in an orthorhombic structural unit cell consisting of 16 rings.\cite{Chen70} Experimentally the sulfur crystal does not have any clean surface cuts and we therefore expect the C-S interface to be quite disordered. In order to reduce the computational cost we have limited our investigation to the interaction between one graphene layer and various sulfur adsorption monolayers. We believe these system to qualitatively capture the graphite-sulfur interface and also simulate a realistic experimental situation where atomic sulfur is deposited on a single graphene sheet.
 %
\subsection{Computational details}
In detail, we have studied the following systems: C$_2$S, C$_8$S, and C$_{18}$S with the S atom in various sites on the graphene surface. The densest S layer, C$_2$S, has an S-S distance of 2.45~\AA\ (theoretical lattice constant for graphene), in comparison with the 2.08 ~\AA\ distance in the S$_8$ ring. The other two systems have the S atoms more and more isolated from each other, simulating a system of isolated S atoms on a graphite surface. STM studies of sulfur electrodeposits on HOPG have shown that sulfur is likely to form islands on the graphite surface in either a C$_2$S configuration or at low concentrations in a $\sqrt{3} \times \sqrt{3}R30^{\circ}$ lattice corresponding to a C$_6$S stoichiometric formula.\cite{Zubimendi96}
In addition, any sulfur-graphite interface can be expected to contain a fraction of S$_8$ rings as well as smaller, reconstructed, rings. We have therefore also studied the interaction of a graphene layer with a full S$_8$ ring, with the ring positioned at different sites on the graphene surface, in order to capture some effects of sulfur molecular interaction with graphite. 
 
All systems were studied employing the first principle density functional theory pseudopotential method \cite{Payne92} implemented in the Vienna Ab-initio Simulation Package (VASP)\cite{Kresse96} with the local density approximation parametrized by Perdew and Zunger \cite{Perdew81} and projector augmented wave pseudopotentials \cite{Kresse99}. The cut-off energy for the plane wave expansion was set to 36.7 Ry and wrap-around errors were avoided in the FFT-grids. Periodic boundary conditions were applied with 22.05~\AA\ of vacuum between periodic graphene layers. In all calculations the carbon atoms were kept fixed whereas the S atoms where allowed to relax in all three spacial directions for the C$_x$S systems but in only the $z$-direction, i.e.\ perpendicular to the graphene plane, for the S$_8$ ring systems. The atomic structures were relaxed until the forces were less than 0.01~eV/\AA. After k-point convergence tests we chose a 11x11x1 $\Gamma$-centered grid for the C$_2$S systems and 7x7x1 grids for the C$_8$S and C$_{18}$S systems. For the C-S$_8$ ring systems we chose a 6x6 graphene unit cell containing 72 carbon atoms and giving a smallest ring-ring distance of 4.05~\AA. Simulations with only S$_8$ in the same unit cell showed that the ring-ring interaction is negligible at this distance. We used a 3x3x1 $\Gamma$-centered grid in this case.
%
\subsection{Results}
\subsubsection{C$_x$S}
For the C$_x$S systems we studied three different high symmetry positions for S on the graphene surface: Site hollow corresponds to S in the middle of the hexagon, site top corresponds to S on top of a C atom, and site bond corresponds to S on top of a C-C bond. We also used several random positions in the $xy$ plane but in no case did these sites have a lower energy than the best high symmetry site.

For the C$_2$S  system the maximal energy difference between the three different sites is very small, 9~meV, with the hollow site slightly favored. STM data for the C$_2$S system has indicated that S is absorbed in the top or bond position.\cite{Zubimendi96} All systems have a distance between the S atom and the graphene surface of 3.3--3.4~\AA, i.e.\ much larger than the S-S distance, implying that the S atoms develop a much weaker bond to the carbon atoms than the S-S bond. 
The C$_8$S and C$_{18}$S systems have artificially much larger S-S distances enforced and as a consequence they develop a stronger C-S bond. For both C$_8$S and C$_{18}$S the top positions are favored, only a few meV below the bond position but $\sim 1$~eV below the hollow position. In the top and bond cases the  S-C surface distance is 1.9--2~\AA. The hollow position retains a larger distance, 2.6--2.7~\AA. These results are in agreement with early quantum chemistry cluster calculations, where the top and bond positions were favored in several configurations of single, double, and triple S atoms configurations.\cite{Vicente96}
If we ignore interface effects, we can qualitatively compare the energies of various systems with the same number of atoms, in order to deduce the energetically favorable S densities. For example, a system with C$_2$S + 3~$\times$~C$_2$ is 2~eV lower in energy than C$_8$S, and C$_2$S + 8~$\times$~C$_2$ is 2.2~eV lower in energy than C$_{18}$S. This indicates that sulfur will prefer to sit on the graphite surface at high densities, and if such densities are not available, sulfur island formation is likely, as also seen experimentally.\cite{Zubimendi96} We also tried to simulate a C$_2$S$_2$ system but this system was found to decompose.

\begin{figure}
\includegraphics[scale = 1]{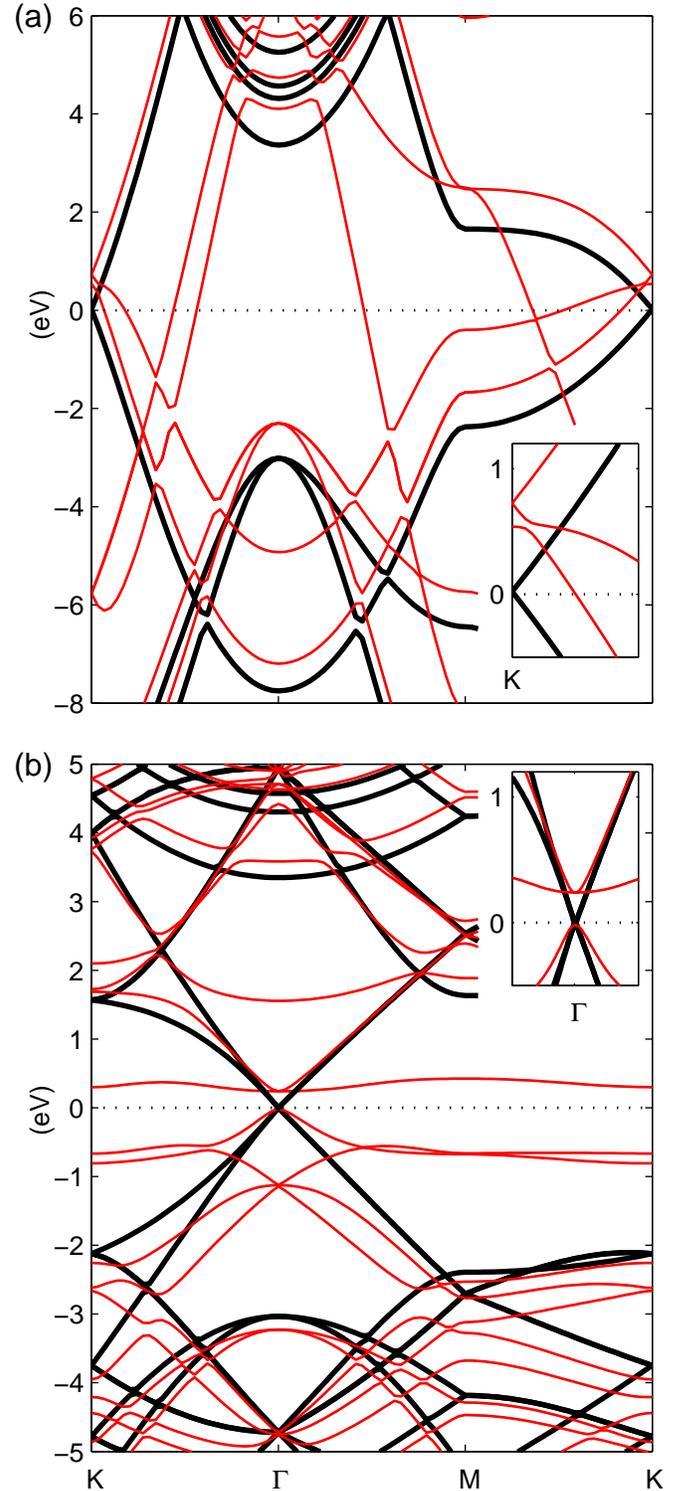}
\caption{\label{fig:CxSband} (Color online) (a) Band structure for C$_2$ (black) and  C$_2$S in the hollow configuration (red). The Fermi level is indicated with a dotted line. Inset shows a zoom-in around the K-point. (b) Band structure for C$_{18}$ (black) and  C$_{18}$S in the top configuration (red). Inset shows a zoom-in around the $\Gamma$-point.}
\end{figure}

Since at high densities sulfur does not form strong bonds to carbon, it is reasonable to assume that the main effect of sulfur on the graphene sheet will only be to induce doping through electron tunneling and not to modify the band structure through chemical bond formation.
This prediction is confirmed in the band structures seen in Figure \ref{fig:CxSband}. 
%
Figure \ref{fig:CxSband}(a) shows the band structure along high symmetry directions in the first Brillouin zone for the C$_2$S system in the hollow structure (red) and pure graphene (black). As seen in (a) and the zoom-in around the K-point (inset) the sulfur in the C$_2$S system shifts the carbon $\pi$-bands upwards by 0.7~eV and induces only a very small hybridization between the carbon and sulfur bands. The upward shift leads to an effective hole doping in the lower $\pi$-band, and most importantly, a finite density of states at the Fermi level for the lower $\pi$-band. 
In order to interpret the physical force behind this upward shift in energy we also calculated the total energy and band structure for a system with the sulfur atoms maximally separated from the graphene layer. In our unit cell that leads to a distance of 11~\AA\ between the sulfur and graphene sheet. The energy difference between this separated structure and the hollow structure proved to be very small, 70~meV. Also, the band structure of the separated system obviously does not show any band mixing but there is still an upward shift of 0.6~eV for the $\pi$-bands. Thus we interpret the C$_2$S hollow system as a consequence of a physisorption process with weak van der Waals bonds between S and C but where the sulfur attracts electrons from the graphene sheet through an electron tunneling process.
The two other systems investigated, top and bond, show similar behavior as the hollow structure, but with a somewhat stronger hybridization between the carbon and sulfur. The density of states at the Fermi level corresponds in these two cases to an upward shift of the $\pi$-band of about 0.4~eV. This difference between the C$_2$S systems can be explained with their actual C-S distance, being 3.6~\AA\ in the hollow case and 3.4~\AA\ in the two other cases.

Figure \ref{fig:CxSband}(b) shows the same results as (a) but for C$_{18}$S in the top (red) position. The top structure has significantly lower energy, 1.4~eV, than the corresponding separated structure where C and S are separated by 11~\AA. This is indicative of a chemisorption process with the formation of a real chemical bond between carbon and sulfur. 
This strong hybridization creates a 0.3~eV band gap in the $\pi$-band at the Fermi level, and we therefore do not have any induced charge transfer in this case. The almost energy degenerate bond site also induces a band gap for the $\pi$-band though a bit smaller in size. The band structure for C$_8$S is qualitatively the same as for C$_{18}$S.

The above energetics and band structures indicate that for sulfur atomic absorption on a graphite surface, high sulfur concentrations  will be preferred, with island formation if necessary, and the main effect on the graphite is a hole doping of the $\pi$-band. For the C$_2$S hollow site system the hole doping is 0.015 holes/C-atom. A graphite layer with this amount doping will have a transition temperature to a $d$-wave superconducting state of $T_c \approx 10$ K for $J/t = 1$, but the value is exponentially sensitive to the value of the coupling constant $J$, for example, $J/t=1.2$ would give $T_c \approx 80$ K. Island formation at a lower macroscopic sulfur concentration would mean that superconductivity is induced locally in the graphite covered by sulfur islands.
\subsubsection{C$_{72}$S$_8$}
For the S$_8$ ring systems we studied five different in-plane positions allowing the $z$-direction coordinate, i.e.\ the height of the ring above the graphite layer, to relax. As a consistency check we also let one of the five configurations relax fully, both C and S atoms. The final atomic structure showed just a 0.02 \AA\  atomic relaxation at most and no energy gain.
The energy difference between all five configurations was small, less than 22 meV, with the lowest energies for the rings that had their sulfur atoms far from any underlying carbon atoms. The closest distance between the graphene surface and the S atoms were in all cases $3.3-3.4$ \AA, indicating a weak bond between C and S. This could have been expected because the sulfur atoms all get a full outer electronic shell by the internal bonding in the S$_8$ ring. Since orthorhombic sulfur is a molecular solid we also expect a minimal electron tunneling between the graphite sheet and the S$_8$ rings. This is confirmed in the band structure calculations that show an unaffected and undoped carbon $\pi$-band for all configurations.
The ring systems are significantly more energetically favorable compared to the atomic sulfur deposition configurations, e.g.\ the C$_{72}$S$_8$ system is  6 eV lower in energy than a system with 8~$\times$~C$_2$S + 28~$\times$~C$_2$ configuration. This is not very surprising considering that the S$_8$ ring is the building block in solid sulfur and it also means that any intact sulfur rings at a graphite-sulfur interface are likely to retain their structure. 
%
%
\section{Intercalated graphite}
Another graphitic system which has, in fact, been known to superconduct for a long time is the intercalated graphites (GIC). They are formed by insertion of metallic atoms in-between the graphite layers. The first discovered superconducting GICs where alkali metal GICs with the composition C$_8$A with A = K, Rb, Cs and with $T_c < 1$ K.\cite{Hannay65}  Using pressure during synthesis it is possible to increase the metallic concentration and in general $T_c$ is found to increase with metal concentration up to $T_c = 5$ K for C$_2$Na.\cite{Belash87} Recently, the intercalated compounds C$_6$Ca and C$_6$Yb were found to be superconducting with $T_c = 11.5$ K and 6.5 K, respectively.\cite{Weller05} These are the highest transition temperatures measured for the GICs as of today.

Intercalating graphite does not significantly change the graphite structure but the intercalated species always donate electrons into the graphite $\pi$-band so the Fermi surface of a GIC can schematically be thought of as $\pi$-band cylinders along HKH and a metallic band, often at the zone center.\cite{Al-Jishi83, Al-Jishi92} However, zone folding will often cause the Fermi surfaces to overlap. For simplicity we will hereafter call the metal band a $s$-band since that is the case in the alkali GICs. 
The charge transfer from metal to graphite is estimated to be be $\sim 0.6$ for C$_8$K\cite{Ohno79} and related compounds and was recently calculated to be 0.32 for C$_6$Ca\cite{Calandra05}. These values give $\delta = 0.08$ and 0.05 respectively, which, if only considering the $\pi$-band, would yield a $T_c$ of several hundred K for a $d$-wave superconducting state. Not only are these values a magnitude or more too large but it has been also found that there is no simple relationship between doping and $T_c$ in the GICs.\cite{Csanyi05} 
However, in most GICs the metal band is only partially occupied and should also be taken into account. While exotic pairing mechanisms have been suggested for especially the high-$T_c$ GICs\cite{Csanyi05} several DFT calculations have recently found large enough electron-phonon couplings to induce the correct $T_c$ in C$_6$Ca and C$_6$Yb.\cite{Calandra05, Mazin05} The phonons are found to mainly couple to the $s$-band since the $\pi$-band will, due to symmetry, not couple to any out-of-plane vibrations, therefore limiting the electron-phonon coupling for the $\pi$-band. It is therefore of interest to consider the possibility to couple the $d$-wave state in the $\pi$-band to a phonon induced $s$-wave superconducting state in the $s$-band.

We have investigated the possibility to couple the two different order parameter states with both band hybridization and interband phonons. Both of these effects are estimated to be small but we are interested in the fundamental possibility to couple the $d$- and $s$-waves to each other, and possibly suppress the $d$-wave state this way.
We consider a mean-field Hamiltonian consisting of only the upper $\pi$-band, since the doping is sufficient to effectively suppress the effect of the lower band and inter-$\pi$-band coupling, and a BCS-like phonon-driven superconducting $s$-band with order parameter $\Delta_s = \frac{1}{N}\sum_{\bf k} \langle s_{{\bf -k}\downarrow} s_{{\bf k}\uparrow} \rangle$. 
In the case of band hybridization, we introduce the additional term $(\Gamma_{\bf k}c^\dagger_{{\bf k}\sigma} s_{{\bf k}\sigma} + {\rm H.c.})$ to the Hamiltonian in order to produce a band mixing. By re-diagonalizing the kinetic energy, i.e.\ calculating the new band structure with the band mixing included, the band operators will change according to $c = \alpha \tilde{c} + \beta\tilde{s}$, with $\alpha\approx 1$ and $\beta \propto \Gamma^2$, and similarly for $s$. Expressed in these new operators the spin-singlet pairing in the original $\pi$-band will induce a $s$-$\pi$ interband pairing as well as an additional pairing mechanism for the $s$-band. The final Hamiltonian can be diagonalized with a double Bogoliubov-Valatin transformation similar to the procedure followed in Section II. The resulting self-consistency equations for $\Delta_{J{\bf a}}$ and $\Delta_s$ have four different solutions of which two are the $d$-wave states for $\Delta_{J{\bf a}}$ with $\Delta_s = 0$ and the final two mixes the $s$-wave state of the spin-singlet pairing with the phonon $s$-wave state. With no coupling between the $d$-wave state and the $s$-wave phonon state and considering only small values of $\Gamma$ we conclude that band hybridization, treated in mean-field theory, will not alter the fact that the $d$-wave solution has a significantly higher $T_c$ than the (coupled) $s$-wave solutions at the considered doping levels.
In the case of interband phonon pairing we used the two-band model introduced by Suhl {\it et al.}\cite{Suhl59} but here with a $k$-dependent coupling in the $\pi$-band. Even here, the $d$-wave solution does not mix with the $s$-wave phonon pairing.
In both cases, the lack of coupling between the two different order parameters can be traced back to the fact that for spherical or 6-fold symmetric bands, the two-fold degenerate $d$-wave solution will have a zero overlap with any functions whose $k$-dependence comes solely from the band structures. In fact, in order to achieve a non-zero $\Delta_s$ even below $T_c$($d$-wave) the six-fold symmetry of the quasi-particle energy has to be broken in order to create non-zero overlap integrals.

The above results indicate that the two-fold degenerate $d$-wave superconducting state in the $\pi$-band will not couple to a phonon-driven superconducting state in the metal $s$-band. With the doping levels relevant for the GICs this means that the unaffected mean-field $d$-wave solution does not agree with experiments.
We can see at least two possible explanations to this. Quite fundamentally, with a partially full $s$-band present, the electronic screening of the on-site repulsion in the $\pi$-band will be significantly enhanced. This should mean that the effective coupling $J$ is drastically reduced and thus no or only a very small effective attraction exists between electrons in the $\pi$-band.
Alternatively, or in addition, there might be important many-body effects that our in the mean-field solution has ignored. Recent band structure calculations of GICs have shown that the Fermi surfaces of the $\pi$-band and the $s$-band is likely to overlap.\cite{Csanyi05,Calandra05} One way to model the effects of such overlap would be to introduce an interband tunneling term of the form $T_{\bf k}s^\dagger_{{\bf k}\sigma}c_{{\bf k}\sigma}$ as in the periodic Anderson impurity model but not treat it as the band mixing term above. It is important to note that such a model for graphite will include a significant kinetic energy term compared to a periodic Anderson impurity model where the impurity lattice is supposed to not have any intrinsic hopping. Colloquially speaking, the coupling to an external band will cause fluctuations of electrons between the two bands which effectively might suppress the spin-singlet interaction in the $\pi$-band. Such an effect might suppress the superconducting correlations in the $\pi$-band to the degree to favor the phonon-driven superconductivity in the GICs instead. It is worth noting that while the atomic sulfur layer in the C$_2$S systems show metallic behavior we do not expect a bulk $s$-band since sulfur is a bulk insulator and thus this effect should not be present in the graphite-sulfur composites.
 
%
%
%
\section{Concluding remarks}
Assuming there are pronounced spin-singlet correlations in graphite we have shown that a $d$-wave TRS breaking superconducting state is possible even for only moderately doped graphite sheets. The intrinsic carrier concentration in graphite due to the interlayer coupling as well as reported electrostatically induced doping in single graphene sheets are however both too small to give a measurable $T_c$. 
%
%
%
So far we have, however, completely ignored the effect of defects and natural adsorption of gaseous molecules in graphitic systems. It has recently been shown that in graphene extended defects such as dislocations, disclinations, edges, and micro-cracks will lead to self-doping where electronic charge is transferred to or from the defect to the bulk.\cite{Peres06} Also, natural chemical doping has been detected on graphene.\cite{Novoselov04} We expect both mechanisms to be more present in graphite samples\cite{Kopelevich06} than in graphene where the play only a minor role. It should then, in principle, be possible to achieve local areas in a graphite sample where the electron or hole doping is large enough to either cause a condensation of the RVB pairs or at least enhance their correlations enough in order to measure a superconductor-like response of the sample. It is natural that such mechanisms are very sensitive to heat treatment as well as aging, as found experimentally.
%
%
%
%
%

In the graphite-sulfur composites the situation is slightly more favorable since atomic sulfur will tend to hole dope graphite. As we have shown it is possible to achieve doping levels of the order of 0.015 holes per C-atom in the $\pi$-band for atomic sulfur deposition on a graphite sheet. This is large enough to induce $d$-wave superconductivity with experimentally measurable transition temperatures within the graphite layer. In the graphite-sulfur composites we anticipate the interface to be disordered with occurrence of both atomic sulfur and sulfur rings. The sulfur rings do not influence the graphite layer to any large extent but the atomic sulfur will cause an electron transfer to the sulfur atoms and thus induce superconductivity. The anticipated disorderedness of the graphite-sulfur interfaces and the variable sizes of the graphite and sulfur domains will make also these systems sensitive to heat treatment and aging as recently demonstrated.\cite{Kopelevich06}

We have so far interpreted the mean-field transition temperature as the phase transition temperature to a superconducting state, as valid in BCS theory. However, with an underlying real-space short-range pairing and, in many cases, small Fermi surfaces in graphitic systems, it is reasonable to question if mean-field theory is able to estimate the superconducting transition temperature and not just to give a temperature scale for onset of incoherent mean-field electron pairing. The thus formed bound-electron pairs might eventually condense but then through a Bose-Einstein condensation (BEC) and then at an often significantly lower temperature. The so-called BCS-BEC crossover problem (for reviews see e.g.\ \onlinecite{Randeriabook95, Loktev01, Chen06}) describes the problem of discerning the appropriate theory for a specific system with effective fermion attraction. 
Pistolesi {\it et al.}\ \cite{Pistolesi94,Pistolesi96} showed that the parameter $k_F \xi$, which measures the coherence length $\xi$ relative to the interparticle distance, is a natural parameter to follow the evolution from BCS to BEC theory. They also showed, by using BCS theory, that it can be approximated using the relationship $T_c/T_F \approx 0.4/(k_F\xi)$, where $T_F$ is the Fermi temperature. This relationship will work for Dirac fermions as well if identifying $k_F$ by $E_F = \hbar v_f k_F$, where $E_F$ is the Fermi energy and $v_f$ is the Fermi-Dirac velocity.
Without going beyond mean-field theory we therefore estimate $k_F\xi = 0.4/(T_c/T_F) \sim 300$ for the C$_2$S system. Following Ref.\ \onlinecite{Pistolesi94}, that places it in an Uemura plot\cite{Uemura91} significantly closer to the classical BCS conductors than the high-$T_c$ exotic superconductors. For the GICs, the steep increase in $T_c$ with only moderate increase in doping level cause them to fall within critical distance of the crossover to BEC theory at $k_F\xi \sim 2\pi$. This estimate strongly questions our mean-field $T_c$ as the true superconducting temperature for the GICs as well as other potential graphitic systems with a high level of doping and thus a high predicted $T_c$. 
In addition, for $d$-wave symmetry the effective electron pair size is always finite and it has been shown that in this case the superconducting bosonic regime is in fact never reached.\cite{Chen99,Chen06} In combination the $d$-wave symmetry and the closeness to a BEC-like theory might strongly influence the value of $T_c$ for highly doped graphite bands, possibly to the extent as to completely suppress superconductivity. Intriguingly, however, it should still be possible to observe consequences of the effective pairing attraction even in these materials. For the GICs, however, this is under the assumption that the large metallic $s$-band does not suppress correlations all-together as argued above.

%
%
%
It might also be worth commenting on the two-dimensionality of our model. In graphite there exists a non-zero interplanar coupling which should make a mean-field transition approximately valid. In the case of interface superconductivity, such as suggested above for the graphite-sulfur composites, the transition should instead be of the Kosterlitz-Thouless (KT) type found in thin films. Here the mean-field transition temperature will give the temperature scale at which the amplitude of the order parameter becomes non-zero whereas phase coherence occurs only at the lower KT transition temperature.

In summary we have demonstrated the possibility to achieve $d$-wave superconductivity in graphite layers at low but non-zero doping levels through an effective favoring of spin-singlet nearest neighbor bonds. The $d$-wave state is TRS breaking, opening the possibility of coexisting ferromagnetic and superconducting correlations. We have also shown by DFT calculations that sulfur can induce high enough hole doping levels in graphite to render the interface in graphite-sulfur composites superconducting through our proposed mechanism. Finally, we have noted the possible failure of mean-field theory to estimate $T_c $ for the GICs and other highly doped graphitic systems. 

%
%
\begin{acknowledgments}
We thank Kyeongjae Cho for providing the computer resources for the DFT calculations 
at the San Diego Supercomputer Center. A.M.B.-S. gratefully acknowledges support from DOE contract DE-AC02-76SF00515.
\end{acknowledgments}


\end{document}